\documentclass{aa}

\usepackage{graphicx}
\usepackage{txfonts}
\usepackage{natbib}
\bibpunct{(}{)}{;}{a}{}{,}

\begin{document}

\title{Simulating planet migration in globally evolving disks.}

\author{A. Crida\inst{1}, A. Morbidelli\inst{1} \and F. Masset\inst{2}}

\titlerunning{Planet migration in globally evolving disks.}

\authorrunning{A. Crida \it et al.}

\offprints{A. Crida}

\institute{Observatoire de la C\^ote d'Azur, B.P. 4229, 06304 Nice Cedex 4, FRANCE\\
           \email{crida@obs-azur.fr}
           \and
           UMR AIM, DSM/DAPNIA/SAp, Orme des Merisiers, CE-Saclay,
              91191 Gif/Yvette Cedex, FRANCE\,;\\
           IA-UNAM, Apartado Postal 70-264, Ciudad Universitaria,
              Mexico City 04510, MEXICO}

\date{Received\,: June 21, 2006 / Accepted\,: August 29, 2006}

\abstract{Numerical simulations of planet-disk interactions are
usually performed with hydro-codes that -- because they consider only
an annulus of the disk, over a 2D grid -- can not take into account
the global evolution of the disk. However, the latter governs
planetary migration of type~II, so that the accuracy of the planetary
evolution can be questioned.}
{To develop an algorithm that models the local planet-disk
interactions together with the global viscous evolution of the disk.}
{We surround the usual 2D grid with a 1D grid ranging over the real
extension of the disk. The 1D and 2D grids are coupled at their common
boundaries via ghost rings, paying particular attention to the fluxes
at the interface, especially the flux of angular momentum carried by
waves. The computation is done in the frame centered on the
center of mass to ensure angular momentum conservation.}
{The global evolution of the disk and the local planet-disk
interactions are both well described and the feedback of one on the
other can be studied with this algorithm, for a negligible additional
computing cost with respect to usual algorithms.}
{}


\maketitle

\section{Introduction}

Planetary formation occurs in disks of gas and dust around
protostars. In particular, giant planets -- whose mass is mostly made
of hydrogen and helium -- must have formed before the dissipation of
the gas disk. Consequently, they must have exerted tidal forces on the
gas and their orbits must have evolved in response to the gas.

In particular, the presence of a planet on a circular orbit leads to
the formation of a spiral density wake in the disk. \citet{GT80}
and \citet{LinPapaloizou1979} showed that, through the
wake, the planet gives angular momentum to the part of the disk
exterior to its orbit, while it takes angular momentum from the
inner part. For small mass planets, \citet{Ward1997} showed that the
net result is a loss of angular momentum for the planet, which makes
its orbit decay on a short timescale. This is usually referred to as
\emph{type~I migration}.

As the planet gives angular momentum to the outer part of the disk, it
repels it outward\,; symmetrically it repels the inner part inward.
If the planet is massive enough (the threshold mass depending on the
disk's viscosity and scale height\,; see \citet{Crida2006} and
references therein) a clear gap opens around the planet's orbit,
effectively splitting the disk into an internal and an external part.
In this situation, the planet gets locked in the middle of the gap,
because both the outer and the inner part of the disk are repellent
for it. Thus, the migration of the planet has to go in parallel with
the migration of the gap \citep{LinPapaloizou1986b,Ward2003}, which
in turn has to follow the viscous evolution of the disk (characterized
by a radial spreading of the disk as the gas is accreted onto the
central star \citep[see][]{LBP74}. This is usually
referred to as \emph{type~II migration}.

Although analytic theories have brought a great deal of understanding
of the fundamentals of planet-disk interactions, it has become
increasingly evident in the last years that numerical simulations are
an essential tool of investigation \citep[see][ for the most recent
review]{Papaloizou-etal-2006}. Numerical simulations, however, are
difficult and time-consuming, and they inevitably involve some
simplifications, which might turn out to produce artifacts. In this
paper we are concerned with the simulation of type~II migration.

As we have seen above, a proper simulation of type~II migration
requires not only the correct calculation of planet-disk interactions
-- which are essentially \emph{local} -- but also of the \emph{global}
evolution of the disk.  Unfortunately, planet migration is typically
simulated with hydro-dynamical codes using a 2 dimensional polar grid
which, for numerical reasons, is truncated at an inner and an outer
radius. This enables to describe well the local interaction of the
planet with the disk but not the global evolution of the disk.

Indeed, the boundary conditions at the extreme rings of the polar grid
can not take into account what happens in the whole disk, outside of
the grid. For instance, the use of \emph{open} boundary conditions
allows the gas to leave the region covered by the grid\,; this makes
the studied disk annulus behave as if it were surrounded by vacuum, so
that it empties very rapidly, which is obviously not realistic. In the
opposite extreme case, boundary conditions which impose that the mean
density on the extreme rings remains constant with time, disable the
accretion and spreading of the gas. Some other prescriptions between
these two extremes may be used, (for instance allowing inflow, or
imposing an outflow given by an analytical model, or setting the flow
on the last ring equal to the viscous flow measured in the neighboring
rings, etc.). However, it is very difficult to adapt these
prescriptions to the changing behavior of the disk with time, in
particular when the disk undergoes perturbation from the planets. In
any case these prescriptions are rather arbitrary and, thus they may
introduce possible artifacts in the planetary evolution.

In this paper we present a novel idea for the correct calculation of
the global evolution of the disk. It consists in surrounding the 2D
polar grid with a 1D radial grid. The 1D grid extends from the real
inner edge of the gas disk (e.g. the X-wind truncation radius at a few
tenths of AU) to the real outer edge (e.g. the photo-dissociation
radius at hundreds of AU). This 1D grid has open boundaries at the
inner and outer edges, and exchanges information with the 2D grid for
the definition of realistic, time-dependent boundary conditions of the
latter. Our algorithm for the interfacing between the 1D and 2D grids
is driven by the requirement that the angular momentum of the global
system (the disk in the 2D section, plus the disk in the 1D section
plus the planet-star system) is conserved.  We will describe it in
detail in Sect.~\ref{sec:coupling}.  First, however, we need to
revisit the algorithm used for modeling the gas evolution and
planet-disk interactions on the 2D grid, to ensure that it also
conserves angular momentum, which is often not the case in standard
implementations. We discuss this issue in Sect.~\ref{sec:cons2D}.
In Sect.~\ref{sec:results}, the results of our new algorithm will be
discussed, in terms of CPU time and robustness with respect to the
positioning of the interfaces between the two grids. Finally in
Sect.~\ref{sec:applics}, we will describe some interesting
astrophysical applications of this new hydrodynamical code.

\section{Conservation of angular momentum in 2 dimension hydrodynamical 
algorithms}

\label{sec:cons2D}

Consider a simulation of the planet-disk interactions, with the disk
represented on a 2D polar grid with open boundary conditions.  A
necessary requirement for the simulation to be correct is that the sum
of the angular momenta of the disk, of the star-planet(s) system and
of the gas outflowing through the boundaries, remains constant over
time.

We have tested if this is the case, using the code FARGO
\citep{FARGO}.

Our simulation accounts for a Jupiter mass planet on an initially
circular orbit. Here and in the rest of the paper we adopt the
following units\,: the solar mass, the initial semi-major axis of the
planet and its orbital frequency, so that the gravitational constant
$G=1$ and an orbit lasts $2\pi$ time units at $r=1$. The grid
used to represent the disk extends from $r=0.25$ to $r=3$ with open
boundaries. It is equally divided in $N_r=165$ elementary rings, and
$N_s=320$ sectors. The disk aspect ratio ($H/r$) is set uniformly
to $5\%$. It is assumed to be constant in time, hence the disk is
locally isothermal. The equation of state used in FARGO is\,:
$P={c_s}^2\Sigma$, with $c_s=H\Omega$ as usual, where $P$ is the
pressure, $\Sigma$ the density, $c_s$ the sound speed, $H$ the disk
scale height, and $\Omega$ the local angular velocity. The gas
kinematic viscosity is $\nu=10^{-5.5}$ in our normalized units (which
corresponds to the viscosity at the location of the planet for
$\alpha=1.25\,10^{-3}$ in a \citet{ShakuraSunyaev1973}
prescription). Its mean density is $\Sigma=3.10^{-4}$, which is a bit
less than the Hayashi Minimal Mass Solar Nebula at Jupiter
\citep{Hayashi1981}\,; the initial density profile can be seen in
Fig.~\ref{fig:profil2D1D} and corresponds to a disk that evolved for
some time under the effect of its own viscosity\,; it can be very well
approximated by $\Sigma(r)=0.000306\,\exp(-r^2/52.8)$.

\begin{figure}
\includegraphics[width=0.7\linewidth,angle=270]{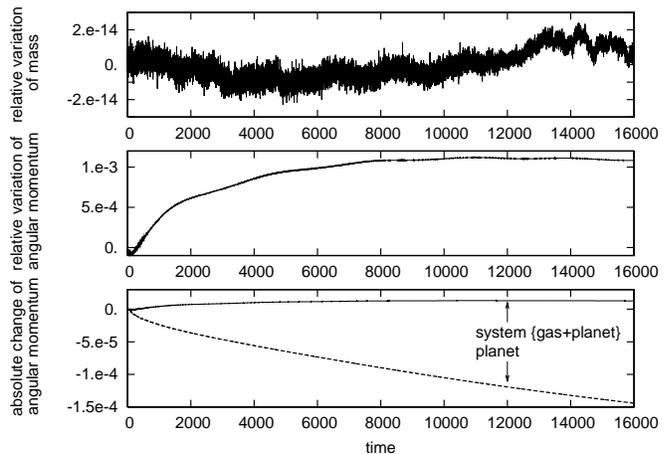}
\caption{Relative variation of mass (top panel) and angular momentum
(middle panel) of the system \{gas + Jupiter + outflow\} with
time. The absolute variation of angular momentum of the system is
compared to that of the planet on the bottom panel (in our units, in
which the initial angular momentum of the planet is $10^{-3}$).}
\label{fig:cons2D}
\end{figure}

The evolution of the total mass and angular momentum of the whole
system (including the outflow, e.g. the cumulated mass and momentum
advected outside of the region spanned by the 2D grid by the gas
outflowing through the boundaries) is presented in
Fig.~\ref{fig:cons2D}. As one sees, while the total mass is conserved
at the level of numerical errors (top panel), the conservation of the
angular momentum is quite poor (middle panel).

The bottom panel shows that the gain of angular momentum of the whole
system (plain line), which is the error of the simulation, amounts to
$10\%$ of what the planet exchanges with the gas in its migration
process (dashed line). Thus we can expect that the migration of the
planet measured in the simulation is correct only at the $10\%$ level.

This shows the need to improve the algorithm in order to achieve a
much better conservation of angular momentum. We describe a procedure
that considerably improves angular momentum conservation below.

\subsection{Choice of reference frame}

\paragraph{\it Need of a frame centered on the center of mass.}
Most, if not all, the published numerical simulations of disk-planet
interaction that we are aware of used a non-inertial frame centered on
the primary \citep{Crida2006, Kley1999, Masset2002, Masset-etal-2006,
Nelson-etal-2000, NelsonBenz2003, Varniere2004, Varniere2005}. This is
done for practical reasons, and because it is thought to be better
adapted to describe the motion of the inner part of the disk.  This
reference frame is not inertial, so that indirect forces are taken
into account for the simulation to be realistic. It is well known that
the presence of these forces destroys the conservation of the angular
momentum measured in the non-inertial frame. In principle however, one
can make at any time a change of coordinates to compute position and
velocities in the inertial frame (centered on the center of mass), and
the angular momentum computed from these coordinates should remain
conserved. This is for instance what happens in N-body codes, when the
simulation is done in heliocentric coordinates (except from truncation
and round-off errors).  This is not the case here. The transport
algorithm for the gas -- in which mass, angular momentum, linear
momentum etc. advect from a cell to its neighbors (sub-step 5 in
Sect.~\ref{subsec:order}) -- imposes the conservation of each of these
quantities.  This is correct for the mass, but not for the momenta,
because they are not supposed to be conserved in the adopted frame. As
a consequence of this imposed, unphysical conservation, the
conservation of the momenta in the inertial frame is corrupted. The
solution out of this problem is the use of the frame centered on the
center of mass throughout the algorithm.

\paragraph{\it Implementation.}
This is not trivial to implement. A first possibility is to suppress
all indirect forces in the algorithm and, whenever the position and
the velocity of the star are needed, to compute them imposing that the
position of the center of mass of the whole system (star, planets and
disk) is at (0,0). This is, for instance, what is usually done in
N-body barycentric simulations. In this case, however, the situation
is more complicated. While in N-body codes all interactions are
treated simultaneously, planet-disk interaction simulators involve a
purely gravitational part and an hydrodynamical part, and treat the
planetary system and the disk separately (see \ref{subsec:order} for
the sequence of integration sub-steps). When the disk is advected, the
star is moved to assure the position of the center of mass, but not
the planet. Thus, at the next step, the planet sees a star in a
different location in phase-space with respect to the end of the
previous step. This has dramatic effect for the orbital stability of
the planet and the overall conservation properties. A second, more
advantageous possibility is to consider the star exactly like a planet
during all the stages of the computation. This ensures a symmetry in
the treatment of the star and of the planet(s). When the gas is
advected, the planetary system (now including the star) is translated
in phase-space to ensure that the center of mass of the whole system
remains motionless at $r=0$. This little correction does not perturb
the relative planetary motion. It corrupts, in principle, the
conservation of the total angular momentum, but the errors introduced
have negligible consequences, as we will test and discuss below.

\paragraph{\it Caveats.}
Because the grid is centered on the center of mass, a good numerical
accuracy can be achieved only if the motion of the gas is approximated
by a rotation around the center of mass. This is the case as long as
the central object is close to the center of mass, namely if the total
mass of the planets is negligible with respect to the one of the star
(which is the common case) or -- in case of a large stellar companion
-- if one considers a distant circumbinary disk. The choice of a frame
centered on the center of mass is obviously not adapted for studying a
disk around a star with a massive companion or a circumplanetary disk.

In the case of an axisymmetric problem (accretion disk with no
planet), the choice of such a frame may lead to numerical
instabilities because small deviations from axial symmetry in the disk
due to numerical errors cause a shift of the star relative to the
center of mass, which in turn can enhance the disk's asymmetries. Thus, if
the planetary system is made of the sole star, we recommend to keep it
fixed at the origin of the frame.

Finally, we remark that in many applications, the equations of motion
are implemented in a rotating frame in order to keep the planet at a
constant position. This can still be done in a frame centered on the
center of mass.

\subsection{Sub-steps sequence}
\label{subsec:order}

As we anticipated above, the integration algorithm treats separately
the pure gravitational part and the hydrodynamical part. To ensure a
good preservation of the conserved quantities, it is necessary to
respect as much as possible the action-reaction principle in the
planet-disk interactions. This requires that the sequence of
integration sub-steps is taken in a specific order.

Here is the sequence that we adopt\,:
\begin{enumerate}
\item The gravitational potential of the planets and star is computed
and stored.
\item The velocities of the planets and star are updated using the
gravitational influence of the disk.
\item The velocities of the gas are changed in each cell according to
the non advective part of the Navier-Stokes equations (the external
forces dues to the gradients of pressure and gravity field and the
viscous stress)\,; the gravitational potential computed at sub-step 1
is used.
\item The planet-star system is advanced under its own gravitational
interactions using a N-body algorithm (specifically a 5th order
Runge-Kutta integrator).
\item The advective part of the Navier-Stokes equations is
performed\,: using the disk velocities computed at sub-step 3, the
mass, angular and radial momentum are advected from each cell to the
neighboring cells. The new velocities of the gas are then computed in
each cell from the new momenta and masses.
\item The conservation of the center of mass is ensured by translating
in space and velocity the center of mass of the planets-star system.
\end{enumerate}
As one sees, in the above algorithm, every time that an action is made
on the planetary system, the equivalent action is immediately applied
on the disk. Moreover, there is no difference in treatment between
planets and star, which is essential for the reasons discussed
above. However, particular care has to be paid in the implementation
of the algorithm to ensure the effective symmetry of these actions\,;
this is discussed in next subsection.

\subsection{Action-reaction symmetry}
\label{subsub:symmetry}

The action-reaction principle has to be perfectly fulfilled in the
computation of sub-steps 2 and 3. Notice that it is not the case, for
instance, if one computes the action of a planet on the disk from the
gradient of its potential on each cell, and the action of the gas on
the planet from the sum of the elementary gravitational forces that
it feels from each cell. In principle, both calculations should be
equivalent, giving total forces of the same strength and opposite
directions. However, if the gradient is computed by finite
differences (typical in grid calculations), this introduces a
difference with respect to the force exerted by the corresponding cell
on the planet.

In order to impose action-reaction symmetry, we proceed as follows.
After sub-step~1, we measure the total change of angular momentum that
the gas will have at sub-step~3 due to the potential of each
planet. Then, in sub-step~2, we compute the change of azimuthal
velocity of each planet according to the change of angular momentum
that it causes to the disk.  This ensures angular momentum
conservation between sub-step~2 and sub-step~3. Similarly, in sub-step
2 the planet's radial velocity change is computed by measuring the
total change of linear momentum in this direction caused to the
gas disk by the same planet.

\subsection{Results and discussion}
\label{sub:res2D}

\begin{figure}
\includegraphics[width=0.7\linewidth,angle=270]{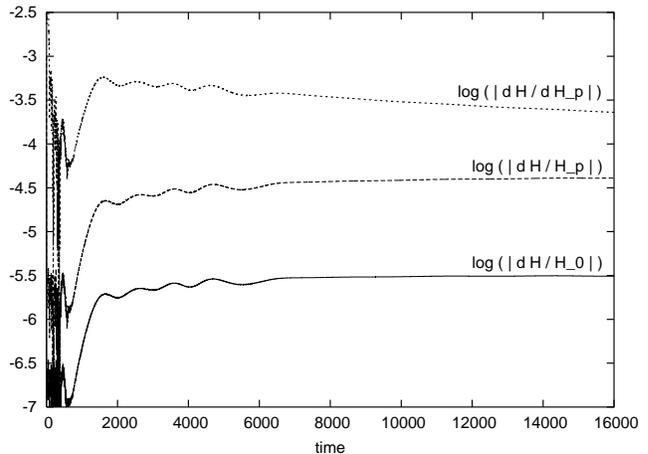}
\caption{Relative angular momentum variation as a function of
time. The variation of angular momentum of the system $\{$ gas +
planet + star + outflow $\}$, $\delta H$, is compared to its initial
value $H_0$ (bottom curve, plain line), to the angular momentum of the
planet $H_p$ (middle curve, long-dashed line), and to the variation of
the latter $\delta H_p$ (top curve, short-dashed line).}
\label{fig:cons2D_good}
\end{figure}

All these precautions allow us to improve greatly the angular momentum
conservation of the whole system. The same simulation of
Fig.~\ref{fig:cons2D} is computed with our modified algorithm. We
denote by $\delta H$ the variation of angular momentum, in
the inertial frame centered on the center of mass, of the whole
system (the sum of the momenta of the gas in the 2D grid, of the
planet-star system, and of the outflow which, we remind, is the
cumulated momentum carried by the gas that flew out of the grid). It
should be zero if the code were perfectly conservative. Thus $\delta
H$ is a measure of the error of the integration scheme. It is compared
with three relevant quantities in Fig.~\ref{fig:cons2D_good}.  The
bottom curve (solid line) corresponds to the logarithm of the relative
variation of angular momentum of the system\,: $\delta H/H_0$, where
$H_0$ is the initial total angular momentum (it corresponds to
the middle panel of Fig.~\ref{fig:cons2D})\,; on a long timescale
(16000 time units, $\gtrsim$ 2500 orbits at $r=1$), this normalized
error does not exceed $10^{-5.5}$. The middle curve (long dashed line)
corresponds to the logarithm of $\delta H/H_p$, where $H_p$ is the
angular momentum of the planet\,: the error in total angular momentum
is about $4.5$ orders of magnitude smaller than the angular momentum
of the planet. Thus, it should not affect its migration. Indeed, the
top curve (short dashed line) shows the logarithm of the ratio between
$\delta H$ and $\delta H_p$, where $\delta H_p$ is the variation of
angular momentum of the planet\,: this ratio is smaller than
$10^{-3}$.

Although highly satisfactory for our purposes, the conservation is
admittedly not perfect, in particular when compared to the mass
conservation (which is in the new simulation as good as in the example
of Fig.~\ref{fig:cons2D}). This is due to two sources of error.

The most obvious one is that the algorithm used for the advancement of
the planetary system, which is in our case a Runge-Kutta algorithm, is
not symplectic. However, we checked that in our simulation, the amount
of angular momentum artificially introduced in the system in this way
is 3 orders of magnitude smaller than the total global error.

The second source of error is in sub-step 6, due to the translation of
the center of mass of the planet-star system. This translation occurs
for two reasons. (i) It is required to compensate the center of mass
motion due to errors introduced by the discretization of the grid and
to the second order (in time) advection of the disk. Thus, the
amplitude of this effect decreases with increasing disk resolution and
decreasing time-step. (ii) Gas continuously leaves the grid through
its inner or outer boundary. This gas is deleted from the simulation,
while its total mass and momentum are recorded as outflow. The problem
is that the outflowing gas is in general not axisymmetric.  Therefore,
its elimination from the simulation causes a shift (in position and
velocity) of the center of mass of the whole system. The shift is
artificial, in the sense that it would not exist if the gas were
modeled with an infinite 2D grid, and it requires, for compensation,
the translation of the planetary system. This issue is a conceptual
one. Thus this source of error cannot be reduced by tuning the grid
resolution or time-step. We think that there is no way around it\,: it
is the price to pay to work with 2D grids that are not as extended as
the physical disk. However, as we have seen in
Fig.~\ref{fig:cons2D_good}, this noneliminable error is so small that
for all practical purposes we can safely ignore it.

\section{Coupling of the 2D grid with a 1D grid}

\label{sec:coupling}

As we discussed in the introduction, for a correct simulation of
type~II migration the conservation of angular momentum is a necessary,
but not sufficient condition. It is also necessary that the global
evolution of the disk is correctly reproduced, which is not the case
if only a portion of the disk (the annulus represented by the 2D grid)
is considered.

For this reason, we consider the whole disk, from its real inner to
outer radius, corresponding respectively to the `X' radius at the
inner edge \citep{Shu2002}, and to a photo-evaporation radius at the
outer edge \citep[see][]{HollenbachAdams2004a,
HollenbachAdams2004b}. Because the representation of the whole disk
with a 2D grid would be computationally prohibitive, outside of the
radial distance range covered by the usual 2D grid we study the
evolution of the extended disk in one dimension, with the aid of a 1D
grid representation. In practise, the 1D grid is made of two parts\,:
one extending from the real inner edge of the disk to the inner edge
of the 2D grid, and the other one extending from the outer edge of the
2D grid to the real outer edge of the disk.

In this section, we first present the equations for the 1D disk. Then
we detail how we couple the 1D and 2D portions of the disk in a way
that conserves mass and angular momentum, and that ensures a proper
simulation of the disk's global evolution.

\subsection{1D equations}
\label{1Deqs}

The Navier-Stokes equation and the advection of the gas are computed
in the 1D grid exactly like in the 2D grid, but with $N_s=1$, so that
each cell can be considered as an elementary ring. Consequently,
there are no azimuthal components for the various gradients, neither
an advective transport phase in the azimuthal direction. The
gravitational potential due to the planets and the star is computed
for the 1D grid as a function of $r$ only\,: the whole planetary system
is considered as a single central mass, the mass of which is the sum
of all the bodies inside the considered radius. Denoting by
$\mathcal{H}$ the Heaviside function equal to $1$ for positive
arguments and to $0$ for negative ones, the gravitational potential
felt by a cell of the 1D grid located at a radius $r$ from the
center of mass reads\,:
$$
\phi^{\rm 1D}(r)=\sum_p -\frac{GM_p}{r}\mathcal{H}(r-r_p)
$$ where the index $p$ goes through the whole planetary system
(including the star). This means that the potential felt is equal
to $-G/r$ multiplied by the mass of all the celestial bodies that lie
inside the ring. Thus the position of the star with respect to the
center of mass has no influence\,; it would be the same if the
1D rings were centered on the star.

We assume that the gas disk in the 1D grid is axisymmetric, so that in
principle there is no torque in the interaction between the
planets and the 1D grid. For more realism, however, one can assume
that each 1D ring feels a gravitational torque due to each planet,
using the formula of \citet{GT80} or \citet{LinPapaloizou1979}\,:
\begin{equation}
\delta T_g(r) \approx 0.4\,\left(\frac{M_p}{M_*}\right)2 {r_p}^3
{\Omega_p}^2 r^{-1} \left(\frac{r_p}{\Delta}\right)^4 \left(2\pi r
  \Sigma\,\delta r\right)
\label{eq:1Dtorque}
\end{equation}
where $\Delta=r-r_p$, $\Omega_p$ is the angular velocity of the
planet, and $\delta r$ is the width of the ring. Of course, for
angular momentum conservation, the opposite of this torque exerted on
the 1D grid has to be exerted on the planet.

Because of the assumption of axial symmetry, what leaves the 1D grid
through its boundaries does not impose any re-adjustment of the
star-planet(s) center of mass. Thus, the conceptual problem discussed
at the end of \ref{sub:res2D} is not relevant here. Consequently, the
evolution on the 1D grid always shows a perfect angular momentum
conservation, at the level of numerical errors.

Despite of this perfect conservation, however, for the results to be
correct it is necessary that the assumption of axial symmetry is a
good approximation for the real disk. Moreover, because
Eq.~(\ref{eq:1Dtorque}) is an approximation, it is necessary that the
planetary torque is small for the error to be small in absolute value.
Both requirements are fulfilled if the interfaces between the 1D and
2D portions of the disk are placed sufficiently far from the
planet(s). We will detail on this issue in Sect.~\ref{sec:results},
with quantitative tests.

\subsection{Ghost rings}

\begin{figure}
\begin{center}
\includegraphics[width=\linewidth]{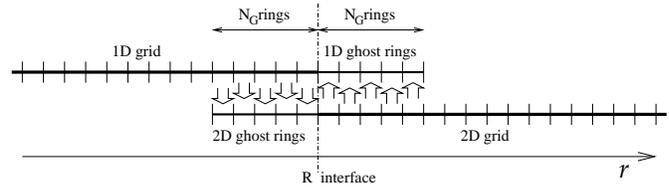}
\caption{Sketch of the coupling via ghost rings at the inner interface
between the 1D grid and the 2D grid.}
\label{fig:coupling_scheme}
\end{center}
\end{figure}

The 1D grid and the 2D grid communicate with each other through a
system of \emph{ghost rings}. This technique is inspired from that
used for the parallelization of a hydro-code on a distributed memory
architecture. We describe it here for the inner interface between the
1D grid and the 2D grid, that is the interface at the inner edge of
the 2D grid\,; the 1D grid is inside, extending from the real inner
edge of the disk (the truncation radius or the surface of the star) to
the 2D grid\,; the 2D grid lies outside this interface, as sketched in
Fig.~\ref{fig:coupling_scheme}.

Outside the interface, a number $N_G$ of 1D rings are added, which
superpose to the first $N_G$ rings of the 2D grid\,; they are the
ghosts for the inner 1D grid (see Fig.~\ref{fig:coupling_scheme}). At
the beginning of every time-step, the density and the velocity in the
1D ghost rings are set as follows. The density of gas ($\Sigma$),
radial momentum ($\Sigma v_r$), and angular momentum ($\Sigma r
v_\theta$) in the ghost ring are set to the azimuthal average of the
respective quantities in the corresponding ring of the 2D
grid. Denoting with the superscript `1D' the quantities in 1D rings
and by $h$ the specific angular momentum ($h=rv_\theta$), this leads
to the following equations\,:
\begin{eqnarray}
\label{eq:Sigma1D}
\Sigma^{\rm 1D} & = & \frac{1}{2\pi r}\int_0^{2\pi}\Sigma \,r{\rm
d}\theta\ ,\\
\label{eq:vrad1D}
\Sigma^{\rm 1D}{v_r}^{\rm 1D} & = & \frac{1}{2\pi r}
\int_0^{2\pi}\Sigma v_r\,r{\rm d}\theta\ ,\\
\label{eq:vtheta1D}
\Sigma^{\rm 1D}h^{\rm 1D} & = & \frac{1}{2\pi r}\int_0^{2\pi}
\Sigma h\,r{\rm d}\theta\ .
\end{eqnarray}
As $\Sigma^{\rm 1D}$ is given by Eq.~(\ref{eq:Sigma1D}), ${v_r}^{\rm
1D}$ and ${v_\theta}^{\rm 1D}$ are easily computed from
Eq.~(\ref{eq:vrad1D}) and (\ref{eq:vtheta1D}).

Then, during the time step, the computation of all stages is performed
in the 1D grid as well as in the ghost rings. The 1D grid `feels' the
ghosts via the Navier-Stokes equations. As the 1D ghost rings have
been filled with quantities inherited from the 2D grid, the 1D grid
behaves as if it felt the 2D grid outside the interface.

Symmetrically, inside the interface, a $N_G$ number of 2D rings are
added\,: the 2D grid ghosts. They are treated as normal 2D rings,
except that at the beginning of every time step, the azimuthal means
of surface density, of radial momentum and angular momentum density
are set equal to the respective values measured in the corresponding
1D rings, as sketched with arrows in
Fig.~\ref{fig:coupling_scheme}. So, the 2D grid effectively `feels'
the 1D grid inside the interface. Filling the 2D ghosts is a bit more
elaborated than in the 1D ghost rings case. Indeed, for a given 2D
ghost ring, one has to proceed in four steps\,: (i) store for each
cell the surface density, the radial momentum, and the angular
momentum\,; (ii) compute the azimuthal means of these quantities\,;
(iii) subtract in each cell, for each of the three considered
quantities, the difference between its azimuthal mean and its value in
the corresponding 1D ring\,; (iv) find the velocity in each cell by
dividing the new momenta by the new surface density.

The minimal number of ghost rings needed, $N_G$, depends on the
numerical scheme\,: it is the number of rings that causally affects a
given ring during a time-step, or equivalently the number of rings on
which the information contained in a ring propagates during a
time-step\,; it is often called the kernel. For instance in FARGO,
$N_G=5$.

\begin{figure}
   \centering
   \includegraphics[width=0.7\linewidth,angle=270]{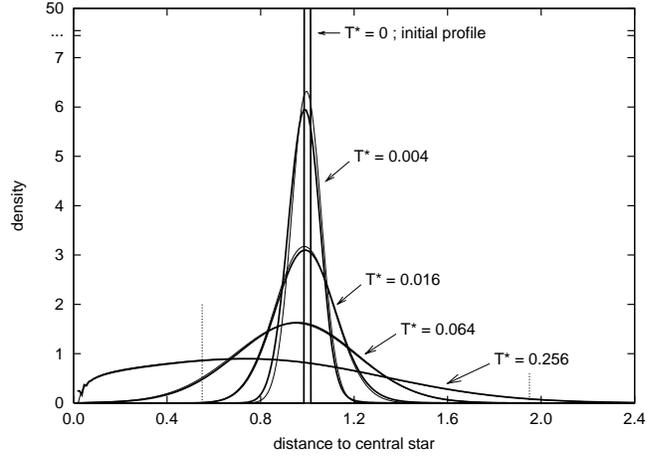}
   \caption{Viscous spreading of a ring. The theoretical curves
   (obtained by numerical solution of Eq.~(14) of \citet{LBP74} with the
   boundary condition $\Sigma(0.02)=0$) are thin, while the
   numerical ones obtained with our code are bold. The vertical dashed
   lines show the position of the interfaces, where the solution provided by
   our code remains perfectly smooth.}
   \label{fig:test_LBP}
\end{figure}

This way of coupling the 1D and 2D grids ensures a smooth transition
for each of the quantities computed in the code. In particular, if
there is no planetary perturbation, the global disk (2D + 1D parts)
behaves exactly as predicted by the \citet{LBP74} equations. This can
be seen in Fig.~\ref{fig:test_LBP}, which corresponds to Fig.~4 in the
aforementioned paper\,: it shows the evolution of the density
distribution at four times, for $\Sigma_{t=0}(r)=\delta(r\!-\!1)$,
with $\delta$ the Dirac distribution. Like in \citet{LBP74}, $T*$
denotes the viscous time\,: $T*=6\nu t$ in our units. The little
discrepancy between the theoretical profile (thin) and the numerical
one (bold) at $T*=0.004$ most likely comes from the fact that the
initial distribution is not exactly a $\delta$ function\,; this little
difference vanishes with time.

\subsection{Fluxes at the interface}

The fluxes of mass and angular momentum through the interfaces have to
be perfectly conservative. What leaves the 2D grid has to enter the
surrounding 1D grid, and vice-versa.

\subsubsection{Mass flux}

\label{subsub:massflux}

The definition of ${v_r}^{\rm 1D}$ in Eq.~(\ref{eq:vrad1D}) makes the
mass flux computed in 1D be the same as the one computed in
2D. Indeed their theoretical expressions write\,:
\begin{eqnarray*}
{F_m}^{\rm 2D} & = & \int_0^{2\pi}\Sigma v_r\,r{\rm d}\theta\\
{F_m}^{\rm 1D} & = & 2\pi r\Sigma^{\rm 1D} {v_r}^{\rm 1D}
\end{eqnarray*}
One can see that Eq.~\ref{eq:vrad1D} is equivalent to ${F_m}^{\rm 1D}
= {F_m}^{\rm 2D}$. However, in a numerical scheme with finite time
steps and discrete grid, the flux is most often not computed so
straightforwardly. Care has to be taken that the numerical expressions
of the flux coincide at the interface between the 1D and the 2D
grid. In staggered mesh codes, the density and the velocity are not
defined at the same locations in the cells of the grid\,; typically,
the density is given at the center, while the radial component of the
velocity is given on the middle point of the inner edge and the
tangential component on the middle point of the left edge \citep[see
for instance Fig.~3 of][]{StoneNorman1993}. Thus, the fluxes are most
often computed as\,:
\begin{eqnarray*}
{{F_m}^{\rm 1D}}_{\rm computed} & = & 2\pi r \Sigma^{{\rm 1D}*}{v_r}^{\rm 1D}\\
{{F_m}^{\rm 2D}}_{\rm computed} & = & \sum_j \Sigma^*[j]v_r[j]\,r\delta\theta
\end{eqnarray*}
where the index $j$ goes through all the cells of the ring,
$\delta\theta$ is the angular size of a cell, $r$ is the radius of the
interface, and $\Sigma^*$ and $\Sigma^{{\rm 1D}*}$ are the values of
the densities $\Sigma$ and $\Sigma^{\rm 1D}$ at the interface after
half a time-step \citep[][ paragraph 4.4]{StoneNorman1993}.

To ensure the mass conservation, we are led to solve the
following equation instead of (\ref{eq:vrad1D}) for the ring next to
the interface\,:
\begin{equation}
{{F_m}^{\rm 1D}}_{\rm computed} = {{F_m}^{\rm 2D}}_{\rm computed}\ .
\label{eq:vrad1D_interface}
\end{equation}
This equation is only slightly different from
Eq.~(\ref{eq:vrad1D}). It is an implicit equation in ${v_r}^{\rm 1D}$,
because $\Sigma^{{\rm 1D}*}$ depends on ${v_r}^{\rm
1D}$. Nevertheless, it can be solved by iteration.

\subsubsection{Angular momentum flux}

\label{subsub:Hflux}

Denoting with a prime sign the non axisymmetric components of
the quantities ($h' = h-h^{\rm 1D}$, $v_r'=v_r-{v_r}^{\rm 1D}$),
using Eq.~(\ref{eq:Sigma1D})-(\ref{eq:vtheta1D})\, the angular
momentum fluxes in 1D and in 2D write:
\begin{eqnarray*}
{F_h}^{\rm 1D} & = & 2\pi r\Sigma^{\rm 1D}h^{\rm 1D}{v_r}^{\rm 1D}\\
{F_h}^{\rm 2D} & = & \int_0^{2\pi}\Sigma h v_r\,r{\rm d}\theta\ 
\ = \ {\ F_h}^{\rm 1D} + \int_0^{2\pi} \Sigma h' v_r'\,r{\rm d}\theta
\end{eqnarray*}
Independently of the numerical scheme and of the modification of
${v_r}^{\rm 1D}$ seen before, it appears that it is impossible for the
1D flux to equal the 2D flux. While ${F_h}^{\rm 1D}$ corresponds to
the angular momentum carried by the gas flowing through the interface,
the term $F_h'=\int_0^{2\pi} \Sigma h' v_r'\,r{\rm d}\theta$
corresponds to a flux of angular momentum that is not due to advection
by the axisymmetric flow\,; it comes from the azimuthal perturbations
of $v_r$ and $v_\theta$, that represents the angular momentum carried
by a wave. The wave-carried propagation of angular momentum is
well-known in planet-disk interactions \citep[see for instance
Appendix C of][]{Crida2006, GolreichNicholson1989,
Takeushi-etal-1996}, but cannot be accounted for by a 1D grid. This is
a conceptual problem.

In fact, we face a degree of freedom problem. With only 3 free
variables ($\Sigma^{\rm 1D}$, ${v_r}^{\rm 1D}$, and ${v_\theta}^{\rm
1D}$), 4 quantities have to be set in the 1D ghosts\,: the mass and
the angular momentum that is in each ring, the mass and angular
momentum fluxes. The first two are set with the prescriptions
(\ref{eq:Sigma1D}) and (\ref{eq:vtheta1D}), which determine the values
of $\Sigma^{\rm 1D}$ and ${v_\theta}^{\rm 1D}$. The third variable,
${v_r}^{\rm 1D}$ is naturally set by
Eq.~(\ref{eq:vrad1D})/Eq.~(\ref{eq:vrad1D_interface}). It appears here
that it is impossible to obtain the correct flux of angular momentum
in the 1D grid ghosts.

A possible solution is that, during the advection phase in the 1D part
of the algorithm, one imposes that the angular momentum flowing
through the interface corresponds to the flux computed in the 2D grid,
${{F_h}^{\rm 2D}}_{\rm computed}$. However, as the flux of angular
momentum carried by a non axisymmetric wave cannot be represented with
1D Navier-Stokes equations, $F_h'$ would be deposited abruptly in the
first ring of the 1D grid. This would lead to the formation of a
spurious gap at the interface, and possibly lead to a numerical
instability.

It has been shown that pressure supported waves travel through the
disk and deposit their angular momentum smoothly, as they get
damped viscously \citep{Takeushi-etal-1996} or through shocks
\citep{GoodmanRafikov2001}. Thus, a better solution for our problem
is to model this wave propagation through the 1D grid. We first
perform advection as usual in each grid and evaluate on the interface
$F_h' = {F_h}^{\rm 2D}-{F_h}^{\rm 1D}$ (or, better, $\delta F_h \equiv
{{F_h}^{\rm 2D}}_{\rm computed}-{{F_h}^{\rm 1D}}_{\rm computed}
\approx F_h'$, which might be slightly different from the former
according to the numerical scheme). Then, over a time-step $\delta t$,
we spread in the 1D grid the amount of angular momentum $\delta
F_h\,\delta t$. A prescription for the deposition of the flux can be
found in \citet{GoodmanRafikov2001} as a function of the distance from
the planet. This could be used when there is only one planet, but not
if there are several planets, because it is difficult to know which
fraction of $\delta F_h$ is due to each planet. Thus, we adopt an
exponential function, because it is scale free so that, once $\delta
F_h$ is known at the interface, its deposition does not depend on the
position of the planet(s) and is a function of the distance from the
interface only. In practice, we assume that, over a time-step $\delta
t$, the angular momentum deposited by the wave in a ring of width
${\rm d}r$ located at a distance $d$ from the interface between the 1D
and 2D grids is\,:
\begin{equation}
\delta h=\delta F_h\,\delta t \,\lambda
\exp(-\frac{d}{\lambda})\,{\rm d}r \ ,
\label{deposit} 
\end{equation}
where $\lambda$ is the damping length-scale of the flux. This
assumption will be supported in Sect.~\ref{subsec:R_interface}, where
the evolution of the disk in the 1D grid is found to be insensitive to
the position of the interface with respect to the planet.

We assume that $\lambda=0.5$ in natural units, which is the value
obtained by fitting with an exponential the decay of the wave-carried
flux in appendix C of \citet{Crida2006}. For simplicity, we assume
that $\lambda$ does not depend on the disk's viscosity and
scale-height (an assumption partially justified in
\citet{GoodmanRafikov2001}).

The deposit of the quantity $\delta h$ of angular
momentum in a ring is simulated by applying a suitable torque, namely
by adding to $v_\theta^{1D}$ the quantity $\delta v_\theta^{1D}=
\delta h/r$, where $r$ is the radial distance of the ring from the
star.
 
Notice that the integral of (\ref{deposit}) from $d=0$ to infinity is
equal to the total angular momentum $\delta F_h\,\delta t$ carried by
the wave at the grid interface. However, the 1D disk is not infinite
in radial extent. Thus, a fraction of the angular momentum carried by
the wave will \emph{not} be deposited in the disk, but will outflow
from the system. This outflowing momentum, as well as the angular
momentum and mass advected through the inner and outer radius of the
1D grid are recorded, in analogy with what was done in
Sect.~\ref{sec:cons2D} for the 2D grid alone.

\subsubsection{Mean viscous torque}

The coupling of the 1D and the 2D grid described in the two
subsections above ensures a smooth, conservative evolution of the
disk. The gas is free to accrete and spread from the star to its
physical outer edge, through the interfaces between the
grids. However, at the interfaces, the shear gives an azimuthal
viscous stress. This appears as a torque exerted on a grid by its
ghosts (and reciprocally). The torque exerted by the part of the disk
inside a given radius $r_0$ on the part of the disk outside $r_0$
reads, from the Navier-Stokes equations\,:
$$ t_\nu(r_0)= r_0\int_0^{2\pi}T_{r\theta}\,r_0{\rm d}\theta\ ,
$$
where $\bar{\bar{T}}=\left(
\begin{array}{cc}
T_{rr} & T_{r\theta}\\
T_{\theta r} & T_{\theta\theta}
\end{array}
\right)$ is the stress tensor, $\bar{\bar{T}}=2\Sigma\nu
\left(\bar{\bar{D}}-(\frac{1}{3}\nabla \vec{v})\bar{\bar{I}}\right)$ ,
with $\bar{\bar{D}}$ the strain tensor and $\bar{\bar{I}}$ the
identity matrix. Thus, $T_{r\theta} = T_{\theta r} = \Sigma\nu
\left(\frac{1}{r}\frac{\partial v_r}{\partial \theta} +
r\frac{\partial (v_\theta/r)}{\partial r}\right)$. In a staggered
mesh scheme, $T_{r\theta}$ is defined at the inner edge of every cell.

For angular momentum conservation, the torque felt by the 1D grid from
its ghosts must equal the one exerted by the 2D grid on its
ghosts. This requires $T_{r\theta}^{\rm 1D}(r_{\rm interface}) =
\frac{1}{N_s}\sum_j T_{r\theta}[j](r_{\rm interface})$, where the
index $j$ goes through the $N_s$ cells of the 2D ring next to the
interface. However, given that the expression of $T_{r\theta}$ implies
the product of velocity gradients by the density, its average is not
equal to the product of the averages, and the required equality is not
necessarily true. Thus, one has to replace the value of
$T_{r\theta}^{\rm 1D}$ at the interface by the average of
$T_{r\theta}$ on the interface in the 2D grid.

\subsection{Results and discussion}

\label{sub:coupling_results}

\begin{figure}
   \centering
   \includegraphics[width=0.7\linewidth,angle=270]{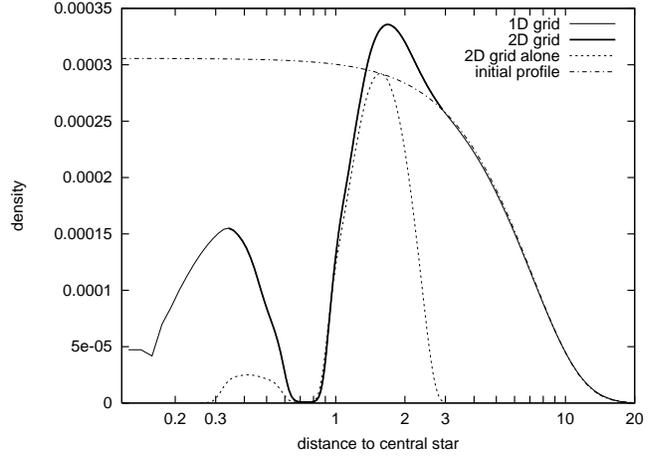}
   \caption{Gas surface density profile after 16000 time units $\sim
   2500$ orbits of the Jupiter mass planet initially placed at
   $r_p=1$. The solid line corresponds to a 2D (bold) and a 1D
   (thin) grid coupled, while the dashed line stands for a 2D
   grid alone. The dot-dashed line shows the initial profile.}
   \label{fig:profil2D1D}
\end{figure}

We consider the final state of the simulation of
Sect.~\ref{sec:cons2D}\,: a Jupiter mass planet initially at $r_p=1$
evolves in a gas disk represented by a 2D grid extending from $r=0.25$
to $3$ for $16000$ time units ($\approx 2500$ orbits). The final
density profile of the gas disk is shown in Fig.~\ref{fig:profil2D1D}
as a dashed line. One can see that the planet has opened a gap and
migrated inward (the planet is located in the middle of the
gap). Moreover, the surface density of the gas over the considered
range is strongly reduced relative to its initial value (dot-dashed
line in Fig.~\ref{fig:profil2D1D}). We then compute another simulation
with the same planetary system and the same 2D grid, but introducing a
1D grid extending from 0.117 to 20 length units (approximately from
0.5 to 100~AU, assuming the unit length equal to 5 AU). The final
profile of the gas distribution obtained in this new simulation is
shown as a solid line in Fig.~\ref{fig:profil2D1D}. Its bold part
corresponds to the 2D portion of the disk.

We remark two important aspects in Fig.~\ref{fig:profil2D1D}. First,
the radial profile of our new solution is clearly smooth, which
indicates that there are no artifacts in the passage of information
from the 2D grid to the 1D grid and vice-versa. The small kink visible
at the inner boundary of the 1D grid ($r\sim 0.15$) is due to the
implementation of the open boundary condition (see
appendix~\ref{Annex:Open})\,; this artifact also appears in
Figs.~\ref{fig:test_LBP}, \ref{fig:profils2D1D_Rinterf}
and~\ref{fig:profils2D1D_Rinterf_nospreading}. The change of sign of
the second order derivative of the radial profile near the outer
interface ($r\sim 3$) is a real feature at the considered time-step,
as we will discuss later (see Fig.~\ref{fig:profils2D1D_interf-ext}
and the related discussion). Second, despite the disk in the new
simulation has also significantly evolved relative to the initial
profile, the new surface density of the disk is very different from
that obtained in the simple 2D simulation (compare the solid and
dashed curves). In particular, in the new simulation the outer part of
the disk has not significantly evolved, which is due to the long
viscous time there, as the latter scales with $r^2/\nu$. Conversely,
in the old simulation, because of the open boundary condition, we
observed a significant disk erosion. This shows the importance of the
boundary conditions for the global evolution of the system.

\begin{figure}
   \centering
   \includegraphics[width=0.7\linewidth,angle=270]{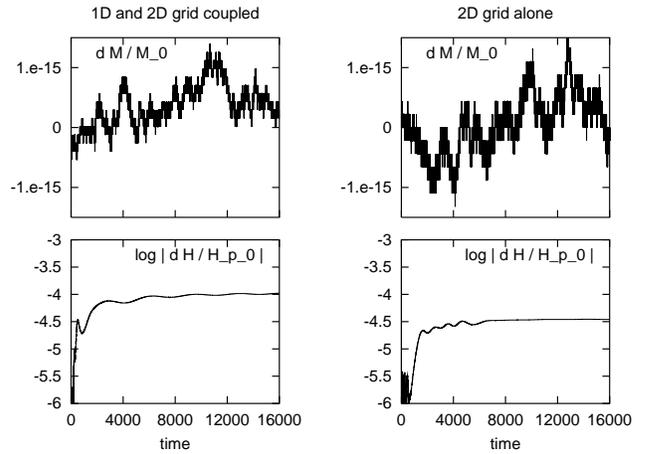}
   \caption{Relative variation of mass and angular momentum in the
   case of a 2D grid with open boundaries and a 2D grid coupled with a
   1D grid.}
   \label{fig:cons2D1Dvs2D}
\end{figure}

The global conservation in the system (namely the evolution of the
mass and angular momentum of the planetary system, plus those of the
gas in the two grids, plus those advected through the inner and outer
radius of the 1D grid and recorded as outflow) is presented in
Fig.~\ref{fig:cons2D1Dvs2D}, and compared to the conservation obtained
in the code using the 2D grid only. One can see that the use of the
extended 1D disk does not change the excellent result obtained in
Sect.~\ref{sec:cons2D}. Thus, the coupling between the two grids that
we described above is correct in terms of conservation
properties. Actually, the error in angular momentum is slightly larger
in the case with the 1D grid because the density at the inner
interface of the 2D grid does not vanish, as can be seen in
Fig.~\ref{fig:profil2D1D}\,; so, the major source of error (the non
axisymmetric outflow from the 2D grid, discussed at the end of
Sect.~\ref{sub:res2D}) does not disappear, while it does for the
simulation using the 2D grid alone.

\section{Performance of the code}

\label{sec:results}

In this section, we check the accuracy of the results, as a function
of the size of the 2D grid, and we discuss the CPU cost of this hybrid
scheme. We show that the coupling of a 1D grid increases by far the
realism of the results of the hydro-code, for a low additional
computation cost.

\subsection{Position of the interfaces with respect to the planet.}

\label{subsec:R_interface}

As already mentioned it in sec.~\ref{1Deqs}, the interface between the
2D and the 1D grids has to be sufficiently far from every planet, so
that the flux of angular momentum carried by the wave at the grid
interface and the azimuthal dependence of the disk surface density are
both small enough. An obvious test of our code is to check the
dependence of the results on the positions of the interfaces between
the two grids.

\begin{figure}
   \centering
   \includegraphics[width=0.7\linewidth,angle=270]{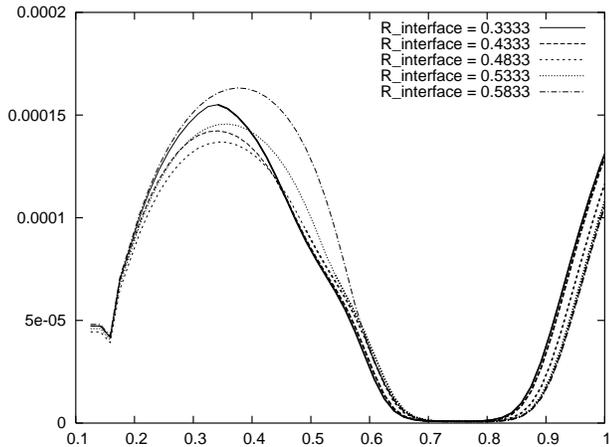}
   \caption{Inner disk profiles at the end of simulations with
   different locations of the inner interface between the 1D and the
   2D grid. The bold part of the profiles corresponds to the 2D grid.}
   \label{fig:profils2D1D_Rinterf}
\end{figure}

We ran a bench of simulations like the one described in
\ref{sub:coupling_results}, with different inner radii for the 2D
grid, but keeping constant the grid resolutions.
Figure~\ref{fig:profils2D1D_Rinterf} shows the density profiles of the
inner disk obtained at the end of the computations ($16000$ time units
$\sim 2500$ orbits). It appears that the differences among the results
are small, even in the cases where the planet is not very far from the
interface ($r_p\approx 0.75$). Only the cases with a transition radius
bigger than $0.5$ show not negligible differences with the three other
runs, which are remarkably similar to each other, in particular for
what concerns the density profile of the inner edge of the gap. The
case with interface at 0.5833 is clearly not accurate, but note that
it is not unrealistic while the 1D grid starts only at 4 Hill radii
from the planet.

\begin{figure}
   \centering
   \includegraphics[width=0.7\linewidth,angle=270]{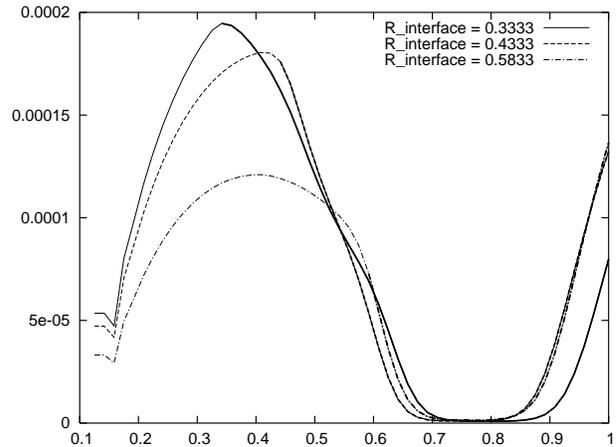}
   \caption{Inner disk profiles at the end of simulations similar to those of 
     Fig.~\ref{fig:profils2D1D_Rinterf}, but without the implementation 
     of the wave damping algorithm (\ref{deposit}). A strong sensitivity on
     $R_{\rm interface}$ appears.}
   \label{fig:profils2D1D_Rinterf_nospreading}
\end{figure}

All cases, however, give a quite consistent representation of the
surface density profile of the inner disk, which is -- on the
contrary -- very different from that obtained by using the 2D grid
alone (compare with the dashed curve in Fig.~\ref{fig:profil2D1D}).

This would \emph{not} be the case if we had not implemented in the 1D
disk the exponential damping of the angular momentum carried by the
density wave launched by the planet. For instance,
Fig.~\ref{fig:profils2D1D_Rinterf_nospreading} shows 3 of the
simulations, recomputed by switching off the calculation of
(\ref{deposit}). One sees a kind of discontinuity at the interface,
where the angular momentum deposition abruptly stops in the disk. This
change implies a modification of the local equilibrium and of the
shape of the density profile. Consequently, the results are strongly
dependent on $R_{\rm interface}$. We note in passing that this also
enlightens the importance for the gap structure of the flux of angular
momentum carried away by the pressure supported wake. This flux is
equivalent to the pressure torque studied in \citet{Crida2006}.

\begin{figure}
   \centering
   \includegraphics[width=0.7\linewidth,angle=270]{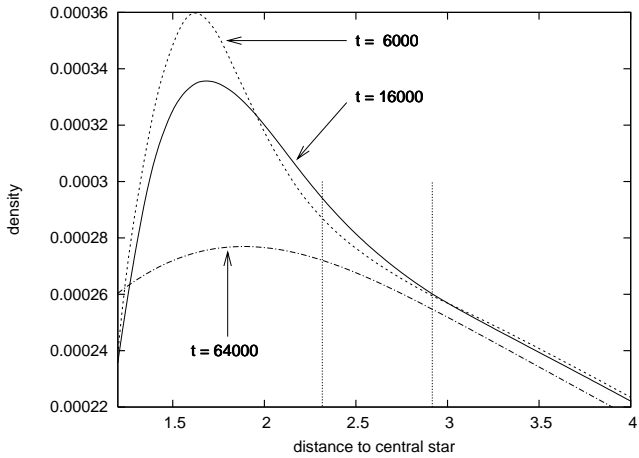}
   \caption{Outer disk profile close to the outer interface ($R_{\rm
   interface}=2.9167$, marked by a vertical line) at 3 different
   times. The vertical line at $r=2.3167$ marks the outer interface
   used in a second simulation, whose results are indistinguishable
   from those plotted in this figure.}
   \label{fig:profils2D1D_interf-ext}
\end{figure}

The outer interface has a much smaller influence on the disk's profile
than the inner one, because it is further from the planet in the
studied case. In Fig.~\ref{fig:profil2D1D}, it appears that the outer
interface corresponds to a change of sign in the second order
derivative of the density profile.
Fig.~\ref{fig:profils2D1D_interf-ext} shows that this is a real
feature, specific to the chosen output time. Indeed, this figure shows
the density profile at different times in the same simulation\,; its
evolution convincingly demonstrates that the second derivative of the
density at the interface ($R_{\rm interface}=2.9167$, right vertical
dashed line) varies with time and can be non-zero. We also repeated
the same simulation, moving the outer interface to $R_{\rm
interface}=2.3167$ (left vertical dashed line). The obtained surface
density profiles overlap exactly with those of
Fig.~\ref{fig:profils2D1D_interf-ext} at the corresponding time, so
that they are not plotted. This test also demonstrates that the
position of the outer interface has a negligible influence on the
global disk evolution.

\begin{figure}
   \centering
   \includegraphics[width=0.7\linewidth,angle=270]{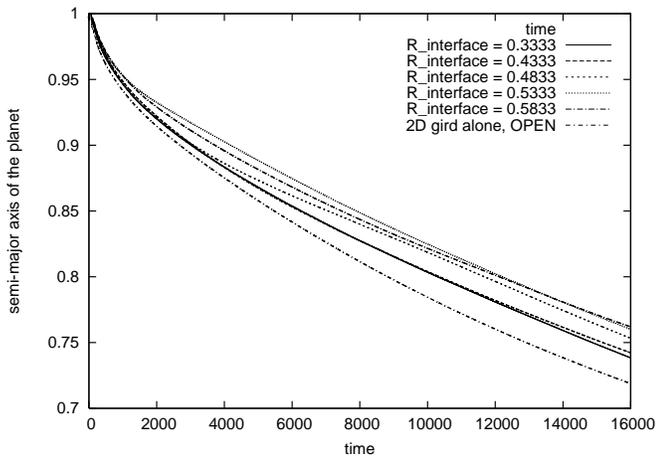}
   \caption{Type~II migration of the Jupiter mass planet in the
   disk. The semi-major axis is plotted as a function of time for all
   the cases already discussed\,: various positions of the inner
   interface between the two grids, and no 1D grid.}
   \label{fig:migrationII}
\end{figure}

The evolution of the disk has a strong influence on the type~II
migration of the planet. Figure~\ref{fig:migrationII} shows the
different migration rates in the simulations of
Fig.~\ref{fig:profils2D1D_Rinterf}, and in the simulation obtained
with the 2D grid alone. Once again, the two first cases ($R_{\rm
interface}=0.3333$ and $0.4333$) are remarkably similar, while the
other ones show a slightly slower inward motion of the
planet. However, the simulations done with the sole 2D grid, gives a
migration that is sensibly faster, as expected due to the
disappearance of the inner disk.

In conclusion, the results presented in this section show that our
method for coupling the 2D and 1D calculations, not only ensures the
conservation of the angular momentum, but also allows a robust
(i.e. weakly dependent on the grid interface position) modeling of the
disk's structure and evolution.

\subsection{Computational cost}

\label{sub:CPU_cost}

In this subsection, we show how the use of a 1D grid allows the study
of the whole disk for a negligible extra cost while the use of an
extended 2D grid would be prohibitive.

From a theoretical complexity point of view, the number of elementary
operations for the computation of a time-step is proportional to
the number of cells of the considered grid. Thus, it is obvious that
the computation of the disk evolution in the 1D grid is negligible
with respect to the computation in the 2D grid.  In addition, one has
to consider the three operations required to couple the grids, which
also need to be done at every time-step.  They are the filling of the
ghost rings, the computation of $\delta F_h$ and its spreading, and
the change of $T_{r\theta}$ in the 1D grid. They require computation
on $N_G$ 2D rings or less. As the size of the ghost area is usually
negligible with respect to the size of the 2D grid, the computational
cost of the coupling is also negligible with respect to the one of a
time-step in the 2D grid.

In the previous paragraph, we studied the number of elementary
operations to be computed during a time-step. In most codes, the length
of the time-step is adapted to the conditions imposed by the required
resolution and the existing perturbation. In general, it is determined
by the Courant-Friedrichs-Lewy (CFL) condition \citep[][ paragraph
4.6]{StoneNorman1993}. The limit most likely comes from the cells with
smallest radii, where the angular velocity is the largest. Indeed, to
avoid numerical instabilities, the time-step length $\delta t$ is set
in order to avoid that information propagates more than one cell over
one time-step\,; thus one has $\delta t\,\Omega < 2\pi/N_s$, where
$\Omega$ is the angular velocity, which is about Keplerian in a gas
disk. Consequently, denoting by $R_{\rm min}$ the inner boundary
radius of the 2D grid, one has $\delta t \propto R_{\rm
min}^{\;3/2}$. This shows that extending the 2D grid toward the star
shortens the time-step and slows down the simulation significantly. The
use of the FARGO algorithm \citep{FARGO} enables one to get rid of the
mean angular velocity and to consider only the perturbed motion and
the shear, but this leads to a similar conclusion, although the
scaling of $\delta t$ with $R_{\rm min}$ is generally different.

The CFL condition in a 1D grid is much less
constraining. Consequently, the addition of a 1D grid inside the inner
edge of the 2D grid has no influence on the time step length, still
determined by what happens in the 2D grid. It is thus important to
increase $R_{\rm interface}$ as much as possible, to speed up the
computation. Some benchmarking confirmed this reasoning (see
appendix~\ref{Annex:Benchmarking} for more detailed results). The
accuracy loss of the computation with the increase of $R_{\rm
interface}$ has been discussed in Subsect.~\ref{subsec:R_interface},
so that an acceptable trade-off can be found.

\section{Astrophysical applications}

\label{sec:applics}

As mentioned in the introduction, the viscous evolution of the
protoplanetary disk is thought to govern type~II migration. Our code,
which has been designed in order to correctly reproduce this evolution,
while resolving the planet-disk interactions, is therefore very
useful for studying any problem related to type~II migration and the
feedback exerted by the presence of planets onto the evolution of the
disk.

We present two problems for which this hybrid scheme is the tool
of choice.

\subsection{Outward migrating planets}

\begin{figure}
   \centering
   \includegraphics[width=0.7\linewidth,angle=270]{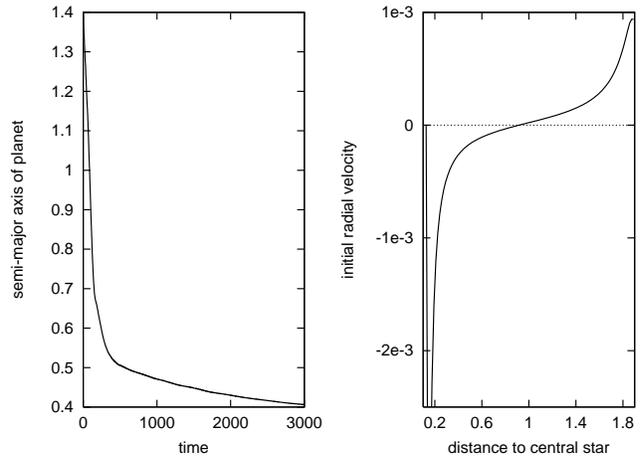}
   \caption{Left panel\,: migration of a Jupiter mass planet initially
   put on a circular orbit at $r_p=1.4$. Right panel\,: Radial
   velocity in the gas disk when the planet is introduced.}
   \label{fig:outwardNO}
\end{figure}

A demonstration of how type~II migration depends on the evolution of
the disk has been provided by \citet{VerasArmitage2004}. A disk
evolving under its own viscosity accretes onto the central star, while
spreading outward under the constraint of angular momentum
conservation. Thus, at any instant in time there is a boundary in the
disk, within which the radial motion of the gas in negative, and
beyond which it is positive -- see the right panel of
Fig.~\ref{fig:outwardNO} and \citet{LBP74}. If a giant planet is
located beyond this boundary it should move outward, as its migration
has to follow the local evolution of the disk \citep[see
also][]{LinPapaloizou1986b}. \citet{VerasArmitage2004} showed
this using a 1D model, where the disk evolved under its own viscosity,
the torque exerted in the planet-disk interaction had the form
(\ref{eq:1Dtorque}), and the effect of waves carrying an angular
momentum flux was neglected.

Using our code, we perform more precise simulations of this
process. The interaction between the planet and the disk is simulated
in 2D, and the effect of waves is taken into account. This implies
that the gap opened by the planet in the disk is less wide and deep
than in \citet{VerasArmitage2004} representation
\citep{Crida2006}. Moreover, the planet also feels a corotation
torque, which is otherwise neglected in the analytic expression of
Eq.~(\ref{eq:1Dtorque}).

Indeed we find that the situation is not as simple as illustrated by
\citet{VerasArmitage2004}. For instance, the left panel of
Fig.~\ref{fig:outwardNO} shows the result of a simulation in which the
planet is initially put at $r_p=1.4$, which is deeply in the outward
spreading zone as shown by the right panel.  Nevertheless, the planet
migrates inward, in what seems to be a runaway type~III migration
\citep{MassetPapaloizou2003}. This is because the spreading disk
forces enough gas to pass through the coorbital region of the planet
so that the planet feels a strong corotation torque \citep{Masset2001}
and decouples from the disk evolution.

We do obtain outward planet migrations, but only in specific
cases. For instance when we place the planet initially outside of the
spreading disk (which is unrealistic), or if we hold it for a
sufficiently large number of orbits, so that it can open a deep gap
that effectively truncates the disk at its inner edge (which, in
practice, places again the planet outside of the spreading disk).

A detailed description of this mechanism and the comprehensive
exploration of the parameter space are beyond the scope of this
paper. However, we think that this example is significant, because it
shows that in reality the evolution of a Jovian mass planet is not an
ideal type~II migration. Because the gap is not extremely clean, the
planet is not fully locked in the disk's evolution. The planet feels
the global motion of the disk (accretion or spreading) but at the same
time it feels also a non-negligible corotation torque. Only a code
like the one that we have developed in this paper can simulate the two
effects correctly and hence allows a quantitative study of the
planet's evolution.

\subsection{Cavity opening}

If the migration of a giant planet is strongly influenced by the
evolution of the disk, the presence of the planet also influences the
evolution of the disk, with a non negligible feedback on its own
migration.

Perhaps the best example is that of the formation of an inner cavity
in the disk. Once a planet has opened a wide and deep gap, the
accretion time of the inner disk onto the star is smaller than the
time scale of planet's migration, because of the negative torque
exerted by the planet \citep{Varniere2005}. This leads to the
disappearance of the inner disk. This in turn enhances the imbalance
between the inner and outer torques felt by the planets, and
accelerates the planet's migration. The presence of cavities are
deduced in some protoplanetary disks, because the spectral energy
distribution (SED) presents a lack of emissivity at the wavelengths
corresponding to hot gas in the vicinity of the star \citep[see for
instance][]{Calvet-etal-2005}. \citet{Rice-etal-2003} suggest the SED
of GM Aur could be explained by a cavity maintained by a two Jupiter
mass planet.

Our algorithm clearly enables a more quantitative analysis of the
cavity opening process than ever done before, for a low computational
cost.  As we have shown in Fig.~\ref{fig:profil2D1D}, when the 1D grid
extends down to the real inner edge of the disk, the evolution of the
surface density of the inner disk is found to be slower than in a
classical code with a truncated 2D grid, which makes the appearance of
a deep cavity more difficult. Therefore, we think that the estimates
of the planet mass required to achieve cavities of given depth, and
the estimates of the lifetimes of these cavities, need to be
revised.  A detailed study of the cavity opening mechanism as a
function of planetary mass, disk viscosity and aspect ratio is in
progress and will be addressed in a forthcoming article.

\section{Conclusion and perspectives}

It is well known that the migration of giant planets (i.e. planets
massive enough to open significant gaps in the disk's density
distribution) is governed -- or at least strongly affected -- by the
global evolution of the disk under its own viscosity.

Usual simulation algorithms solve the hydrodynamical equations over a
2 dimensional polar grid that is truncated at an inner and outer
radius to keep the computing time within reasonable limits. Thus, they
cannot reproduce correctly the evolution of the disk, and no more the
planet's migration.

In this paper, we have shown how the use of a 1D grid surrounding a
classical 2D grid allows us to simulate the global evolution of the
disk, while resolving the local planet-disk interactions with the
accuracy of the usual algorithms. Coupling the two grids via a system
of ghost rings, with special attention paid to the conservation of
the angular momentum, leads to a smooth evolution of the disk from its
inner radius (the truncation radius or the surface of the star) to its
outer edge, several hundreds of AU away. This increases by far the
accuracy of the simulation with essentially no additional CPU cost.

Consequently, this algorithm is a tool of choice to properly simulate
type~II migration, and related problems. More generally, it also
enables the study of the effect of the presence of planets on the
disk's global evolution, which in turn affects the migration of the
planets in a feedback effect.

In our simulations, we observe that the disks slowly disappear through
the open boundaries of the 1D grid. However, disks are believed to
disappear rapidly after only a few million years, under the action of
photo-evaporation. This phenomenon could easily be introduced in our
code via a simplified prescription consisting in removing a fraction
of the gas in each cell, depending on time and location. This will
open the possibility of simulating the evolution of planets in a
globally evolving gas disk, over the disk's lifetime.

\section*{Acknowledgments}

This work was partially supported by a funding from the French
Specific Action for Numerical Simulations in Astrophysics
(A.S.S.N.A.). The authors also wish to thank Hal Levison for
instructive discussion, and an anonymous referee for his interesting
remarks and careful reading.

\appendix

\section{Open boundary condition}

\label{Annex:Open}

The ``open'' boundary condition, that allows outflow from the grid but
not inflow is implemented as follows at the inner edge of the 1D grid.

In the innermost ring (number 0), the radial velocity is always 0 and
the density is the one of the neighboring ring. In the latter (ring
number 1), the radial velocity is set equal to the one of the
following ring (number 2) if and only if it corresponds to outflow
from the grid (negative radial velocity), and to 0 otherwise. This
gives\,:
\begin{itemize}
\item $\Sigma(0)$ is set to $\Sigma(1)$.\\
\item Whenever $v_r(2)\geqslant 0$, $v_r(1)$ is set to $0$.\\
\item Whenever $v_r(2)<0$, $v_r(1)$ is set to $v_r(2)$.
\end{itemize}

So, the first three rings are used for this computation, which
explains the artifact observed in Figs.~\ref{fig:test_LBP},
\ref{fig:profil2D1D}, \ref{fig:profils2D1D_Rinterf}, and
\ref{fig:profils2D1D_Rinterf_nospreading}.

\section{Benchmarking}

\label{Annex:Benchmarking}

We ran 3 simulations of a Jupiter mass planet initially placed at
$r_p=1$, evolving in a gas disk for 1000 time units on an Intel Xenon
2.66 GHz processor. The results, summed up in
table~\ref{table:benchmark} confirm our reasoning of
Sect.~\ref{sub:CPU_cost}.
\begin{table}[h]
\caption{Benchmarking results}
\label{table:benchmark}
\centering
\begin{tabular}{ccccr}
\hline\hline
\multicolumn{3}{c}{parameters of the 2D grid} & 1D grid & CPU time\\
extension & $N_r$ & $N_s$ & extension & of process\\
\hline
0.35 - 3.0     & 165 & 320 & none & 6167 s.\\
0.35 - 3.0     & 165 & 320 & \ 0.1167 - 20 & 6008 s.\\
0.1167 - 5.0 & 294 & 320 & none & 99494 s.\\
\hline
\end{tabular}
\end{table}

First, we see that with or without the 1D grid, the computing cost is
about the same.

Second, it is obvious that the computing time is not directly
proportional to the number of cells, and is strongly increased by the
higher shear in the vicinity of the star, otherwise the third
simulation would not have been more that twice as long as the first
one.

Third, the second simulation was slightly shorter than the first
one because with the 1D grid, the density at the boundaries of
the 2D grid does not tend to zero, the profile is smoother and
consequently the CFL condition is less constraining there.

\bibliographystyle{aa}
\bibliography{crida}

\end{document}